\documentclass{jpsj3}
\usepackage{graphicx,color}

\def\diff{\mathrm d}
\newcommand{\etal}{{\it et al.}\ }

\title{Slab Thickness Dependence of Rashba Splitting on Au(111) Surface: First-Principles and Model Analyses}
\author{Taichi Kosugi$^{1,2}$, Takashi Miyake$^{1,3}$, and Shoji Ishibashi$^{1,3}$}

\inst{
$^1$ Nanosystem Research Institute (NRI) ``RICS'', AIST, Umezono, Tsukuba 305-8568, Japan \\
$^2$ Department of Physics, University of Tokyo, Hongo, Tokyo 113-0033, Japan \\
$^3$ Japan Science and Technology Agency, CREST, Honcho, Kawaguchi, Saitama 332-0012, Japan
}

\abst{
We study the dependence of the spin splitting on the number $N$ of atomic layers,
using first-principles calculation for Au(111) surface.
When the slab of the atomic layers is sufficiently thick, the lower split state has a minimum 
away from $\bar{\Gamma}$, which is known as the Rashba effect. 
As the number of layers decreases, the minimum approaches $\bar{\Gamma}$, and 
it is located at $\bar{\Gamma}$ for $N \leq 14$. 
This crossover is analyzed in detail using two models: 
a tight-binding model and a bilayer nearly-free-electron model.
It is demonstrated that the features of surface band dispersion are clearly understood
as the result of the competition between the interference of the surface states on both sides and the spin-orbit interaction.
}

\kword {gold surface, Rashba splitting, first-principles calculation, electronic structure, spin-orbit interaction}

\begin{document}
\maketitle

\section{Introduction}

Recent progress in spintronics has been increasingly evoking interests in the role of spin-orbit interaction 
in condensed-matter physics. 
Spin-orbit interaction induces many intriguing phenomena, such as the spin-Hall effect, topological insulator 
and magnetoelectric effect. 
At surfaces, it causes spin splittings even in nonmagnetic materials.  
This effect was pointed out by Rashba~\cite{bib:Rashba}. 
He analyzed a two-dimensional electron with a momentum $\boldsymbol{k}$ moving freely on the $xy$ plane under an electric field along the $z$ direction:
\begin{gather}
	H_{\mathrm{R}} (\boldsymbol{k}) = \frac{k^2}{2m} + \alpha \boldsymbol{\sigma} \cdot ( \boldsymbol{e}_z \times \boldsymbol{k} ) 
	\label{Rashba_H}
	,
\end{gather}
where $\boldsymbol{\sigma}$ are the Pauli matrices.
The second term originates from spin-orbit interaction. 
The constant $\alpha$ in the isotropic Rashba Hamiltonian, eq. (\ref{Rashba_H}), is proportional to the gradient of the potential
along the $z$ direction: $\alpha = \frac{1}{4 m^2 c^2} \frac{\diff V}{\diff z}$.
The dispersion relation of this Hamiltonian has two branches for the eigenvalues
\begin{gather}
	E_\pm (\boldsymbol{k}) = \frac{ k^2}{2m} \pm \alpha k
	\label{eigenvalues_monol}
	.
\end{gather}
The two eigenstates for nonzero $\boldsymbol{k}$ have opposite spin polarization.
The spin splitting is explained by the lack of inversion symmetry. 
Kramers' theorem states that in a crystal invariant under time reversal operation $-i \sigma_y K$,
where $K$ is the complex conjugation operator,
two Bloch states having opposite wave vectors and opposite spins are degenerate: $E_{\boldsymbol{k} \uparrow} = E_{-\boldsymbol{k} \downarrow} $.
Independently of this, a system having inversion symmetry satisfies $E_{\boldsymbol{k} \sigma} = E_{-\boldsymbol{k} \sigma}$ for $\sigma = \uparrow, \downarrow$.
Hence in a crystal with both time reversal and inversion symmetry,
$E_{\boldsymbol{k} \uparrow} = E_{\boldsymbol{k} \downarrow}$
must be satisfied, that is, spin splitting is prohibited.
On a surface,
inversion symmetry is broken and spin splitting can occur.

Experimental studies on the Rashba effects reported so far are not only for
surfaces of metals [e.g., Au(111)~\cite{bib:1621,bib:1622}, Ag(111)~\cite{bib:1622}, and Bi(111)~\cite{bib:1629}]
but also for surface alloys [e.g., Ag/Au(111)~\cite{bib:1712,bib:1628,bib:1713}, Bi/Ag(111)~\cite{bib:1396}, and Sb/Ag(111)~\cite{bib:1630},]
including the binary surface alloy Bi$_x$Pb$_{1-x}$Ag(111)~\cite{bib:1626}
and semiconductor heterostructures~\cite{bib:1623,bib:1627},
focusing particularly on how the magnitude of the splitting could be tuned by varying the element and formula.
Among these, the Au(111) surface is a well studied system.

LaShell \etal\cite{bib:1621} observed Rashba splitting, 110 meV at the Fermi level, of the $L$-gap surface states on Au(111) via an angle resolved photoemission spectroscopy (ARPES) experiment,
which is much greater than the zero-field spin splitting for a two-dimensional electron gas, which is on the order of meV in semiconductor heterostructures.\cite{bib:1623,bib:1624}
Their results have been theoretically confirmed using a tight-binding model\cite{bib:862} and first-principles electronic structure calculations.\cite{bib:1622,bib:1236,bib:1424,bib:1237}
The anisotropic Rashba splitting on Au(110) has also been studied by Simon \etal\cite{bib:1523} with the $\boldsymbol{k} \cdot \boldsymbol{p}$ perturbation theory by introducing an anisotropic Rashba term $H_{\mathrm{R}}^{\mathrm{anis}} (\boldsymbol{k}) = \lambda_x k_x \sigma_x + \lambda_y k_y \sigma_y$.
Petersen and Hedeg\r{a}rd\cite{bib:862} pointed out that, in a monolayer tight-binding model for Au(111) 
(where the system has spatial inversion symmetry),
the constant of the Rashba Hamiltonian, eq. (\ref{Rashba_H}), must include the effect not only of the inherent intraatomic spin-orbit interaction 
but also of the transfer integrals $\gamma \equiv \langle p_z (\boldsymbol{R}) |V| p_n (\boldsymbol{R} + \boldsymbol{x}) \rangle (n=x,y)$ between $p$ orbitals deformed by the surface potential for the reproduction of the splitting.
Nagano \etal\cite{bib:1237} demonstrated more quantitatively with first-principles calculations that
the spin-orbit interaction occurs only in the vicinity of each atom\cite{bib:1382} and
that the asymmetric shape of the squared wave function $|\psi(z)|^2$ along the normal of the surface 
is crucial for the large spin splitting.
The asymmetry comes from the mixing of atomic orbitals of different parity,
and the energy splitting is proportional to the integral of the potential gradient multiplied by $|\psi(z)|^2$,
of surface states on Au(111).
It is in contrast with Ag(111) and Sb(111), for which $|\psi(z)|^2$'s are less asymmetric.
The importance of asymmetric feature of surface wave functions was also pointed out for Bi/BaTiO$_3$(001).\cite{bib:1588}

So far, the Rashba effect has been studied for solid surfaces.
However, because of the recent progress in nanoscale fabrication technology,
thin films on substrates are becoming target materials of researches on systems where spin-orbit interactions play important roles.
The possibility of tuning the thickness of an overlayer on a substrate induces interests in the thickness dependence of the Rashba splitting\cite{bib:1712,bib:1628,bib:1713}
from a practical viewpoint since the controllability of the Rashba splitting is desirable for the development of spintronic devices.
In the present work, we investigate the magnitude of the spin splitting as a function 
of the number of layers using both first-principles and model calculations.
The present work also provides technical insights for an electronic structure calculation using a finite slab.
A sufficiently thick slab is quite often used, mimicking a semi-infinite surface system
in an ordinary electronic structure calculation, because it can be treated as a three-dimensional periodic system
when it is separated from its duplicated image by vacuum regions.
Electronic structure calculations for semi-infinite systems without using a slab reported so far are few
due to the complicated formalism and the technically difficult implementation of such calculations\cite{bib:1714,bib:1549}.
We find that, when the slab is thick enough, the energy dispersion of the surface state is well described 
by eq. (\ref{eigenvalues_monol}).
The eigenvalue of the lower branch decreases as the wave vector moves away from $\bar{\Gamma}$. 
We find that, as the slab gets thinner, the energy gap at $\bar{\Gamma}$ becomes larger,
and a crossover takes place.
The energy splitting is the Rashba splitting when the slab is thick, 
whereas it is characterized by hybridization between the surface states of both sides of the slab when it is thin.

\section{Computational Details}

First-principles electronic structure calculation is based on the density functional theory (DFT). 
We adopt the projector augmented-wave (PAW) method\cite{bib:PAW} 
using the Quantum MAterials Simulator (QMAS) package\cite{bib:QMAS} 
within the local-density approximation (LDA).\cite{bib:LDA}  
We implemented fully relativistic calculation using two-component pseudo Bloch wave functions on QMAS. 
The total energy of the system is calculated as a functional of the $2 \times 2$ density matrix defined as
\begin{gather}
	\rho_{\sigma \sigma'} (\boldsymbol{r}) = \sum_{n, \boldsymbol{k}}^{\mathrm{occ.}}
		\psi_{n \boldsymbol{k} \sigma} (\boldsymbol{r})^*
		\psi_{n \boldsymbol{k} \sigma'} (\boldsymbol{r}),
\end{gather}
where $\sigma, \sigma' = \alpha, \beta$ are spin indices. 
It is in contrast to spin-independent nonrelativistic or scalar relativistic DFT calculations,
in which charge density is the fundamental quantity.
The formulation of fully relativistic electronic structure calculations using two-component wave functions has been proposed for
norm-conserving pseudopotentials\cite{bib:1214}, ultrasoft pseudopotentials\cite{bib:1215,bib:1216}, and PAW formalisms\cite{bib:1504} in detail.
In a fully relativistic calculation, noncollinear magnetism and spin-orbit interaction can be naturally introduced.

We solve the Dirac equation as follows.
Under a central electrostatic potential, a four-component energy eigenfunction solution of the Dirac equation is of the form
\begin{gather}
	\psi_{j j_z \kappa} (\boldsymbol{r}) = \frac{1}{r}
	\begin{pmatrix}
		f_{j \kappa}(r) \mathcal{Y}_{j l_{\mathrm{A}}}^{j_z} (\theta, \phi) \\
		i g_{j \kappa}(r) \mathcal{Y}_{j l_{\mathrm{B}}}^{j_z} (\theta, \phi) 
	\end{pmatrix}
	.
	\label{psi_jjzk}
\end{gather}
$\psi_{j j_z \kappa}$ is not an eigenstate of $L^2$, but a simultaneous eigenstate of the operators $H, J^2, J_z, S^2$, and $\mathcal{K}$,
whose eigenvalues are $E, j(j+1), j_z, 3/4$, and $\kappa$, respectively.
$\boldsymbol{J} \equiv \boldsymbol{L} + \boldsymbol{S}$ is the total angular momentum operator.
For the definition and properties of $\mathcal{K}$ in detail, readers are referred to, 
e.g., Ref. 30.
$\kappa$ can take only $j+1/2$ or $-(j+1/2)$.
For $\kappa = \pm (j+1/2)$, $l_{\mathrm{A}} = j \pm 1/2$ and $l_{\mathrm{B}} = j \mp 1/2$.
$\mathcal{Y}_{jl}^{j_z}$ is the spinor spherical harmonics\cite{bib:1042} defined as
\begin{gather}
	\mathcal{Y}_{jl}^{j_z} =
		\begin{pmatrix}
 				\sqrt{\frac{l \pm j_z + 1/2}{2l+1}} Y_{l j_z-1/2} \\
 			\pm \sqrt{\frac{l \mp j_z + 1/2}{2l+1}} Y_{l j_z+1/2}
	\end{pmatrix}
\end{gather}
for $j = l \pm 1/2$,
which is a simultaneous eigenstate of the operators $J^2, J_z, L^2$, and $S^2$.
$f_{j \kappa}$ and $g_{j \kappa}$ are the solutions of the following system of differential equations:\cite{bib:1213}
\begin{gather}
	\Bigg( \frac{\diff}{\diff r} + \frac{\kappa}{r} \Bigg) f_{j \kappa} - \frac{1}{c} (E - V + 2c^2) g_{j \kappa} = 0, \nonumber \\
	\Bigg( \frac{\diff}{\diff r} - \frac{\kappa}{r} \Bigg) g_{j \kappa} + \frac{1}{c} (E - V) f_{j \kappa} = 0.
	\label{Dirac_eq_rad}
\end{gather}
In ordinary condensed matter physics, the upper two components are much larger than the lower two components in eq. (\ref{psi_jjzk}).
We therefore drop the lower components, and $\psi_{j j_z \kappa}$ becomes two-component and now an eigenfunction of $L^2$.
Thus the elimination of $g_{j \kappa}$ from eq. (\ref{Dirac_eq_rad}) leads to the single differential equation to be solved for the construction of a fully relativistic potential:
\begin{gather}
	\Bigg[ \frac{\diff^2}{\diff r^2} - \frac{l(l+1)}{r^2} + \frac{1}{2M(r)c^2} \frac{\diff V}{\diff r} \Bigg( \frac{\diff}{\diff r} + \frac{\kappa}{r} \Bigg) + 2M(r) (E - V) \Bigg] f_{j \kappa} = 0
	,
	\label{radial_eq}
\end{gather}
where we denote $l_{\mathrm{A}}$ by $l$ and
$M(r) = 1 + (E - V)/2c^2$.
The term involving $\kappa$ originates in the spin-orbit interaction and causes $j$-splitting of the energy spectrum.
Since eq. (\ref{radial_eq}) is independent of $j_z$, it has solutions degenerate in $2j+1 = 2l+2$ and $2l$ for $j = l + 1/2 (\kappa = - l - 1)$ and $j = l - 1/2  (\kappa = l)$, respectively, for fixed $l$.
$j$-averaged $\kappa$ is hence
\begin{gather}
	\langle \kappa \rangle
	= (-l -1) \cdot \frac{2l + 2}{4l + 2}
	+ l \cdot \frac{2l}{4l + 2}
	= -1
	.
\end{gather}
If we replace $\kappa$ with $\langle \kappa \rangle$ in eq. (\ref{radial_eq}), the scalar relativistic equation is obtained, whose energy eigenvalues depend only on $l$.
By introducing $\kappa(\lambda) = -1 -\lambda l$ and $-1 + \lambda (l+1)$ for $j = l + 1/2 $ and $j = l - 1/2$ for common $l$, respectively,
we can continuously move from the scalar relativistic ($\lambda = 0$) equation to the fully relativistic ($\lambda = 1$) equation
by varying the strength $\lambda$ of the spin-orbit interaction.

In the present study pseudo wave functions are expanded in plane waves 
with an energy cutoff of 30 Ry for $6 \times 6$ $k$-points for the surface Brillouin zone (see Fig. \ref{Fig_band}(a)),
which give sufficiently converged results for the purpose of the present study.
The slab is constructed by stacking Au atomic layers. 
The lattice constant is fixed to the experimental value 4.078 \AA {} of bulk Au\cite{bib:1000_38}, 
and the slab is separated from its duplicated image in the neighboring unit cell by a vacuum region of 15 \AA {} width.
The atomic positions are not relaxed.

\section{Results and Discussion}

\subsection{First-principles calculation}

\begin{figure}[htbp]
\begin{center}
\includegraphics[keepaspectratio,width=8cm]{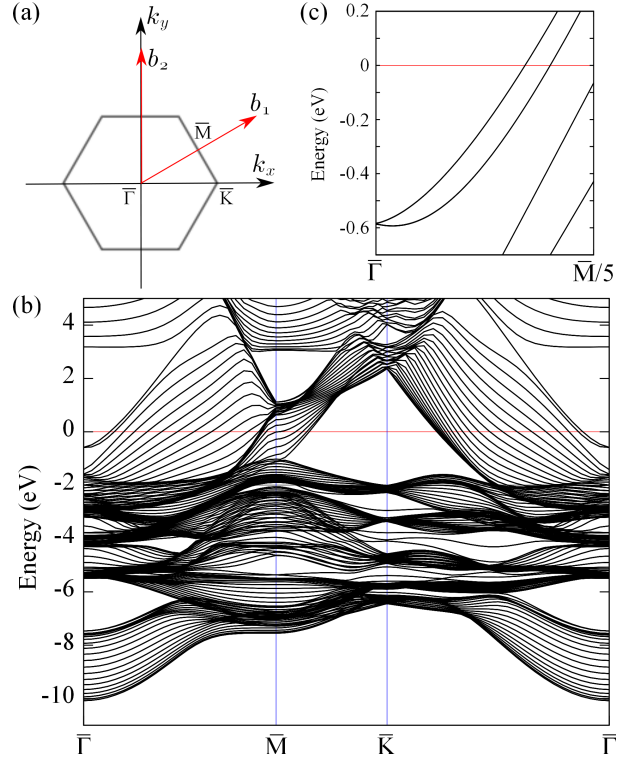}
\end{center}
\caption{
(Color online)
(a) Surface Brillouin zone of Au(111) and primitive reciprocal lattice vectors.
(b) Fully relativistic electronic band structure of a 22-layer slab.
The origin of energy is set to the Fermi level.
(c) is a blowup of (b), showing the vicinity of the spin-split $L$-gap surface states at $\bar{\Gamma}$.
}
\label{Fig_band}
\end{figure}

Figure \ref{Fig_band}(b) shows the fully relativistic electronic band structure of a $22$-layer slab.
There is a bunch of states in the energy region from $-8$ to $-1$ eV. They have a strong $5d$ character. 
The bands having $6s$ and $6p$ orbital characters cross the $5d$ bands. 
Hybridization between them leads to the $sd$-derived surface states near $-7.5$ eV
and the $sp$-derived surface states\cite{bib:1424,bib:1237} near $-0.5$ eV at $\bar{\Gamma}$.
The latter correspond to the experimentally observed\cite{bib:1621} $L$-gap surface states with the spin splitting. 
The experimental and calculated Fermi wave vectors for the inner and outer $L$-gap surface states are shown in Table \ref{Table_splitting}.
Our values are in good agreement with the experimental values.

\begin{table}[h]
\begin{center}
\caption{Experimental and calculated Fermi wave vectors (in \AA$^{-1}$) for the inner and outer $L$-gap surface states.}
\label{Table_splitting}
\begin{tabular}{cccc}
\hline\hline
 &  & $k_{\mathrm{F}}^{\mathrm{in}}$ & $k_{\mathrm{F}}^{\mathrm{out}}$ \\
\hline
Exp. & LaShell {\it et al.}\cite{bib:1621} & $0.153$ & $0.176$ \\
 & Reinert {\it et al.}\cite{bib:1664} & $0.167$ & $0.192$ \\
 & Nicolay {\it et al.}\cite{bib:1622} & $0.172$ & $0.197$ \\
Calc. & Henk {\it et al.}\cite{bib:1236} & $0.149$ & $0.172$ \\
 & Mazzarello {\it et al.}\cite{bib:1424} & $0.159$ & $0.191$ \\
 & Present Work & $0.172$ & $0.201$ \\
\hline\hline
\end{tabular}
\end{center}
\end{table}

\begin{figure}[htbp]
\begin{center}
\includegraphics[keepaspectratio,width=8cm]{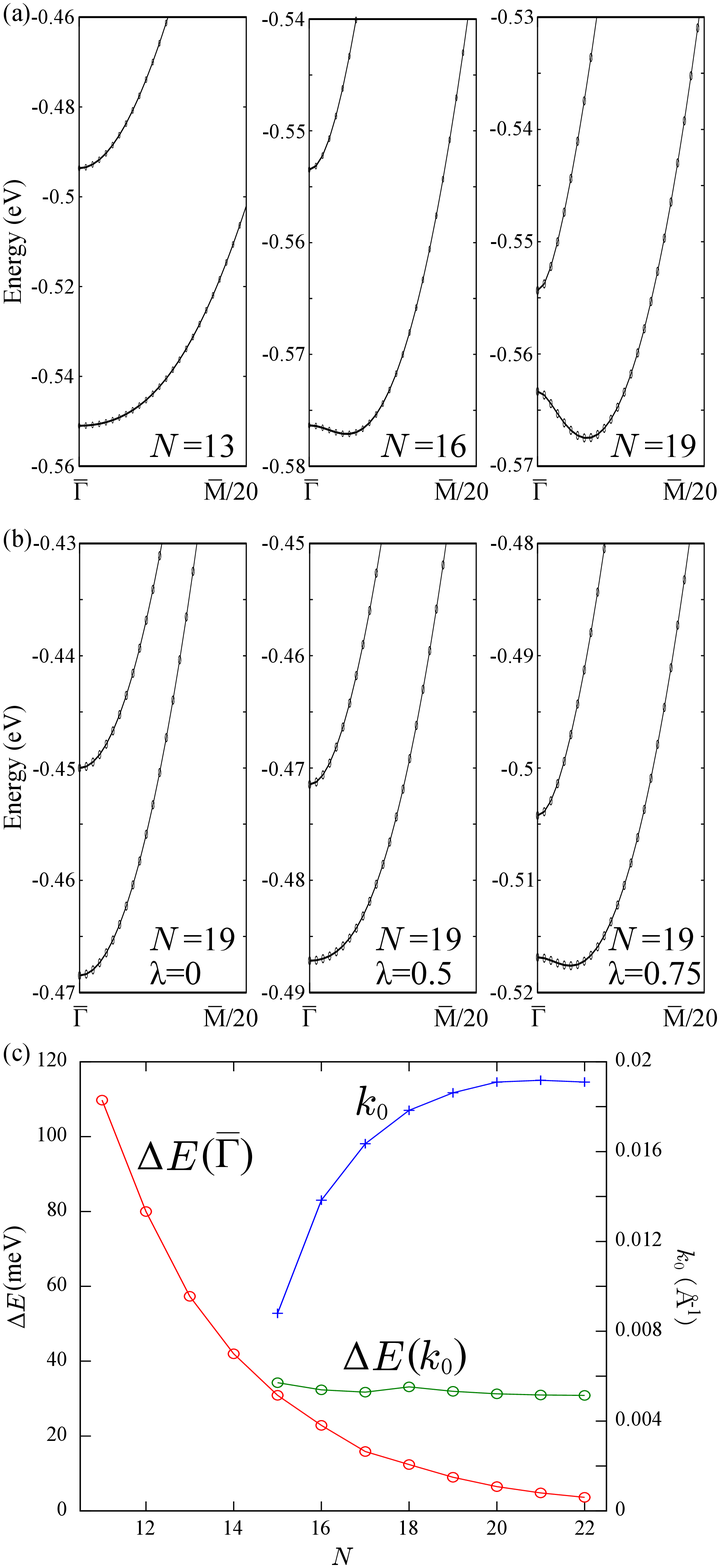}
\end{center}
\caption{
(Color online)
(a) Fully relativistic band structures in the vicinity of the $L$-gap surface states at $\bar{\Gamma}$ for $N = 13, 16$, and $19$.
The origins of energy are set to the respective Fermi levels.
(b) shows the band structure for $N = 19$ with varying the strength $\lambda$ of the spin-orbit interaction.
(c) plots $k_0$ and energy gaps at $\bar{\Gamma}$ and $k_0$ as functions of $N$.
$k_0$ is the finite wave vector, for which the outer surface states take a minimum band energy.
}
\label{Fig_band2}
\end{figure}

We then changed the number of layers and 
carefully examined the features of the band structures of the $L$-gap surface states in the vicinity of $\bar{\Gamma}$. 
Figure \ref{Fig_band2}(a) shows the band dispersion of the $L$-gap surface states at $\bar{\Gamma}$ for $N = 13, 16$, and $19$.
There are two branches, each of which is twofold-degenerate.
(There are four states in total: two degrees of freedom from spin, and the other two from the number of surfaces.)
The calculated energy band of the inner surface states was found to be a monotonically increasing curve 
as a functions of wave vector for all the $N$'s.
It was found, however, that for $N = 16$ and 18, the band takes a minimum value at a nonzero wave vector $k_0$, 
while for $N = 13$ the outer surface state band has a minimum at $\bar{\Gamma}$.

In order to look into this behavior, 
we artificially changed the strength $\lambda$ of the spin-orbit interaction and 
observed the variation of the band structure. 
The result for $N = 19$ is shown in Fig. \ref{Fig_band2}(b).
We can see that 
the bands of the outer surface states reach their minimum only at $\bar{\Gamma}$ for $\lambda = 0$ and $0.5$,
while the minimum is observed off $\bar{\Gamma}$ for $\lambda = 0.75$ and $1$.
This clearly indicates that the spin-orbit interaction is crucial for this behavior. 

We performed systematic calculations for the number of atomic layers $N = 11$-$22$, and found that 
the minimum is away from $\bar{\Gamma}$ for $N \geq 15$.
Figure \ref{Fig_band2}(c) shows a plot of $k_0$ and the energy gaps $\Delta E$ at both $\bar{\Gamma}$ and $k_0$ as functions of $N$.
It is seen that $\Delta E(\bar{\Gamma})$ rapidly decreases as $N$ is increased,
while the variation of $\Delta E(k_0)$ is much smaller 
despite the outward movement of $k_0$ away from $\bar{\Gamma}$.

In the following two subsections, we analyze these results in more detail using 
model calculations. 
It is elucidated from a bilayer model that 
the difference in the features of these bands, which has the critical number of layers, $15$, 
comes from the competition of the strengths of the spin-orbit interaction and the interference 
between the surface states on both surfaces. 

\subsection{Tight-binding model}

Before proceeding to the analysis of a bilayer model,
we examine here the interference of $p$-derived surface states on the uppermost and lowermost layers via a tractable spin-independent tight-binding calculation by varying the number of layers of a slab.

Within the tight-binding formalism,
the Hamiltonian matrix for a wave vector $\boldsymbol{k} = k_1 \boldsymbol{b}_1 + k_2 \boldsymbol{b}_2$ lying on the surface Brillouin zone (see Fig. \ref{Fig_band}(a)) is given by
\begin{gather}
	H_{ij}(\boldsymbol{k}) = \sum_{\boldsymbol{R}} e^{i \boldsymbol{k} \cdot \boldsymbol{R}} t_{i \boldsymbol{0} j \boldsymbol{R}},
\end{gather}
where the sum runs over the lattice points corresponding to the surface unit cell
and $t_{i \boldsymbol{0} j \boldsymbol{R}}$ is the transfer integral between
the orbital $i$ in the home unit cell and the orbital $j$ in the unit cell at $\boldsymbol{R}$.
The orbital indices represent $p_x, p_y$, and $p_z$ orbitals in the present study.
We take into account only the transfers between nearest-neighboring atoms,
$t_p$ and $t_s$, which are often conventionally denoted by $(pp \pi)$ and $(pp \sigma)$, respectively.
The explicit expressions for the intralayer Hamiltonian matrix in bulk, $H_{0}$, are thus easily calculated:
\begin{gather}
	H_{0 xx}(\boldsymbol{k}) = \frac{3}{2} ( c_2 + c_{12}) t_p + \frac{1}{2} ( 4 c_1 + c_2  + c_{12} ) t_s \nonumber \\
	H_{0 yy}(\boldsymbol{k}) = \frac{1}{2} (4 c_1 + c_2 + c_{12}) t_p + \frac{3}{2} ( c_2 + c_{12}) t_s \nonumber \\
	H_{0 zz}(\boldsymbol{k}) = 2 ( c_1 + c_2 + c_{12}) t_p \nonumber \\
	H_{0 xy}(\boldsymbol{k}) = H_{0 yx}(\boldsymbol{k}) = \frac{\sqrt{3}}{2} ( -c_2 + c_{12}) (-t_p + t_s) \nonumber \\
	H_{0 yz}(\boldsymbol{k}) = H_{0 zy}(\boldsymbol{k}) = 
	H_{0 zx}(\boldsymbol{k}) = H_{0 xz}(\boldsymbol{k}) = 0
	,
\end{gather}
where $c_i \equiv 2 \cos 2 \pi k_i (i=1,2)$ and $c_{12} \equiv 2 \cos 2 \pi (k_1+k_2)$.
Those for the interlayer Hamiltonian matrix in bulk, $H_{1}$, are also calculated:
\begin{gather}
	H_{1 xx}(\boldsymbol{k}) = \frac{1}{4}(3 + 3 e_1 + 4 e_{12}) t_p + \frac{1}{4}(1 + e_1) t_s \nonumber \\
	H_{1 yy}(\boldsymbol{k}) = \frac{1}{12}(11 + 11 e_1 + 8 e_{12}) t_p + \frac{1}{12}(1 + e_1 + 4 e_{12}) t_s \nonumber \\
	H_{1 zz}(\boldsymbol{k}) = \frac{1}{3}(1 + e_1 + e_{12}) ( t_p + 2 t_s) \nonumber \\
	H_{1 xy}(\boldsymbol{k}) = H_{1 yx}(\boldsymbol{k}) = \frac{\sqrt{3}}{12}(-1 + e_1) (t_p - t_s) \nonumber \\
	H_{1 yz}(\boldsymbol{k}) = H_{1 zy}(\boldsymbol{k}) = \frac{\sqrt{2}}{6}(1 + e_1 - 2 e_{12}) ( t_p - t_s) \nonumber \\
	H_{1 zx}(\boldsymbol{k}) = H_{1 xz}(\boldsymbol{k}) = \frac{\sqrt{6}}{6}(1 - e_1) ( t_p - t_s)
	,
\end{gather}
where $e_1 \equiv e^{-i 2 \pi k_1}$ and $e_{12} \equiv e^{-i 2 \pi (k_1 + k_2)}$.
We have set the $p$ orbital energies to zero.

We incorporate the surface perturbation on the orbital energies by adding a constant $\varepsilon$,
which is described by the surface effect Hamiltonian
$H_{\mathrm{SE}} \equiv \textrm{diag} (\varepsilon, \varepsilon, \varepsilon)$.
The $3N \times 3N$ Hamiltonian matrix for a $N$-layer slab is thus given by
\begin{gather}
	H(\boldsymbol{k}) =
	\begin{pmatrix}
		H_0 + H_{\mathrm{SE}} & H_1 &  &  &  \\
		H_1^\dagger & H_0 & H_1 & & \\
		&  H_1^\dagger & H_0 & H_1 & \\
		&  &  \ddots &  &  \\
		&  &  H_1^\dagger & H_0 & H_1 \\
		&  &  &  H_1^\dagger & H_0 + H_{\mathrm{SE}} \\
	\end{pmatrix}
	,
\end{gather}
The band dispersion of the slab is obtained by solving the time-independent Schr\"odinger equation $H(\boldsymbol{k}) \boldsymbol{c}_{\boldsymbol{k}} = E_{\boldsymbol{k}} \boldsymbol{c}_{\boldsymbol{k}}$,
where $\boldsymbol{c}_{\boldsymbol{k}}$ is the $3N$-dimensional column eigenvector and $E_{\boldsymbol{k}}$ is its corresponding energy eigenvalue.
The bulk band structure is obtained as a continuum of real $E$ such that real $\theta$ exists satisfying $\det ( H_1^\dagger e^{-i \theta} + H_0 - E + H_1 e^{i \theta} ) = 0$.

We do not incorporate in $H_{\mathrm{SE}}$ the transfer integrals $\gamma$ between $p$ orbitals deformed 
by the surface potential, which were introduced by Petersen and Hedeg\r{a}rd\cite{bib:862} for a monolayer system. 
The reason for this is as follows. 
We study multilayer systems. The translational symmetry is inevitably broken at the surface
and thus the perturbed transfer integrals are not necessary for surface states.
Furthermore, the matrix elements coming from $\gamma$ vanish at $\bar{\Gamma}$
and hence the following discussion would be unchanged even if they were incorporated.

\begin{figure}[htbp]
\begin{center}
\includegraphics[keepaspectratio,width=8cm]{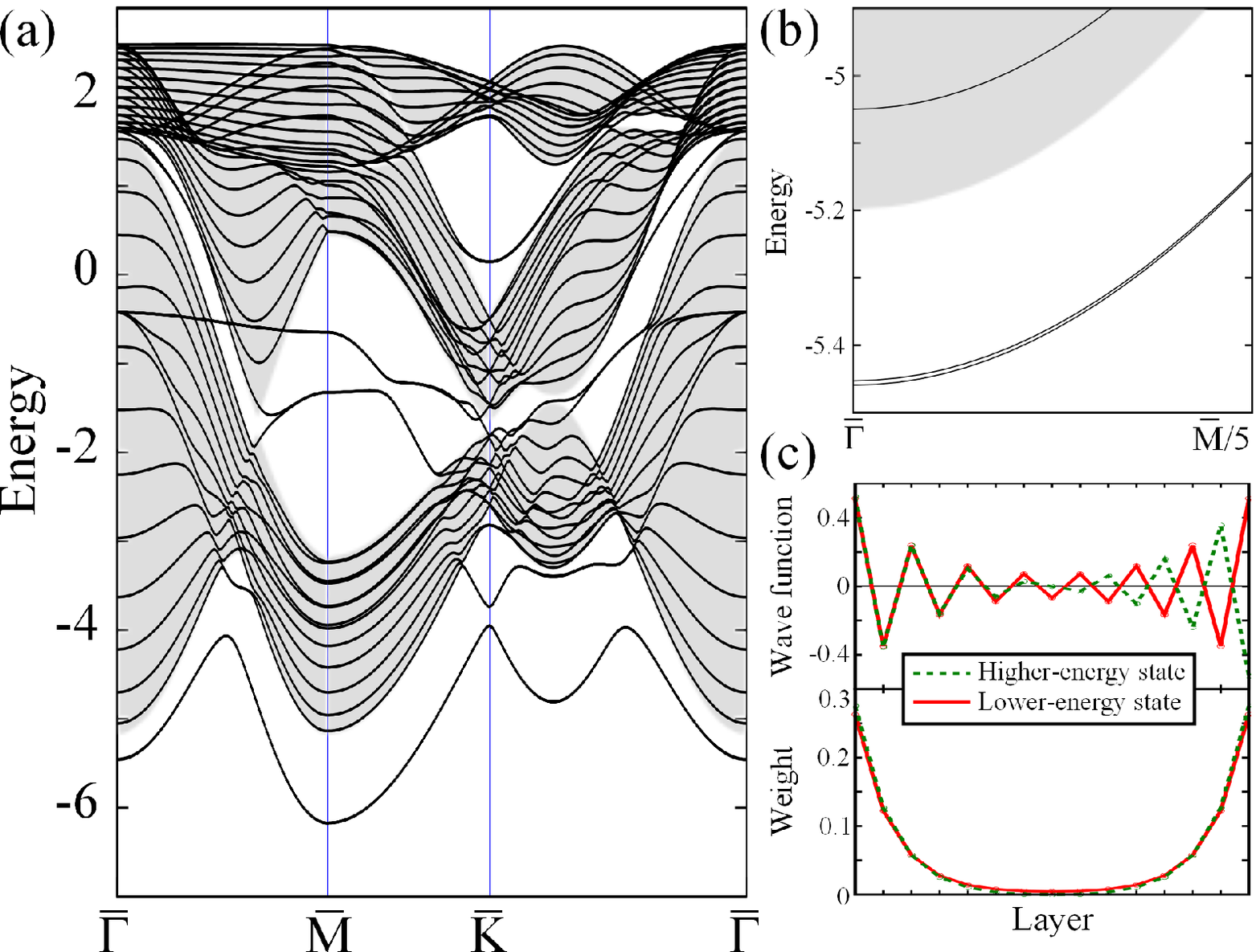}
\end{center}
\caption{
(Color online)
(a) Tight-binding band structure of a 15-layer slab.
The shaded region is the bulk band continuum.
Energies are in units of $t_s$.
(b) is a blowup of (a), showing the vicinity of two purely $p_z$-derived surface states at $\bar{\Gamma}$.
The wave functions and weights of these surface states are shown in (c).
}
\label{Fig_tb}
\end{figure}

Let us inquire into the surface states at $\bar{\Gamma}$.
We rearrange the order of the bases as
$\{ |p_x^1 \rangle, \dots, |p_x^N \rangle, 
|p_y^1 \rangle, \dots, |p_y^N \rangle, 
|p_z^1 \rangle, \dots, |p_z^N \rangle \}$,
where $| p_i^l \rangle$ is the Bloch sum of the $p_i$ orbital on the $l$-th layer for the wave vector $\boldsymbol{k}$.
The Hamiltonian matrix of the slab at $\bar{\Gamma}$ then looks block-diagonal:
\begin{gather}
	H(\bar{\Gamma}) =
	\begin{pmatrix}
		H^\parallel & & \\
		 & H^\parallel & \\
		&  & H^\perp \\
	\end{pmatrix}
	,
\end{gather}
where we have defined $N \times N$ tridiagonal matrices as
\begin{gather}
	H^i \equiv
	\begin{pmatrix}
		t^i_0 + \varepsilon & t^i_1  &  &  & \\
		t^i_1 & t^i_0 & t^i_1  &  &  \\
		 & t^i_1 & t^i_0 & t^i_1  &  \\
		 &  & \ddots &   &  \\
		& & & t^i_1 & t^i_0 + \varepsilon \\
	\end{pmatrix}
\end{gather}
for $i=\parallel, \perp$ and the transfer parameters as
$t_0^\parallel \equiv 3 t_p + 3 t_s, t_0^\perp \equiv 6 t_p, t_1^\parallel \equiv 5 t_p/2 + t_s/2$,
and $t_1^\perp \equiv t_p + 2 t_s$.
Two $H^\parallel$ in $H(\bar{\Gamma})$ correspond to the twofold energy spectra.
One is $p_x$-derived and the other is $p_y$-derived.
$H^\perp$ gives the $p_z$-derived spectrum.
$H^i$ is of the same form as the tight-binding Hamiltonian matrix for a finite one-dimensional chain only with nearest-neighbor transfers between $N$ sites.
Such a system has been analyzed in detail by Davison and Grindlay.\cite{bib:1581}
This one-dimensional system admits two surface states when $|z| > 1 (z \equiv \varepsilon/t_1^i)$.
The eigenvectors corresponding to the surface states are, in the limit of $N \to \infty$,
arbitrary linear combinations of degenerate localized states given by
\begin{gather}
	c_{\mathrm{L} n} = A_N (\mathrm{sgn} \, z)^n e^{-\mu (n - 1)} \\
	c_{\mathrm{R} n} = A_N (\mathrm{sgn} \, z)^n e^{\mu (n - N)} \\
	\mu \equiv \ln |z|
	, \quad
	A_N \equiv e^{\mu (N-1)/2} \sqrt{\frac{\sinh \mu}{\sinh \mu N}}
	.
\end{gather}
$\boldsymbol{c}_{\mathrm{L}}$ and $\boldsymbol{c}_{\mathrm{R}}$ are the solutions decaying from the first and $N$-th sites into the bulk, respectively,
which have the common eigenvalue $E = t_0^i + (\mathrm{sgn} \, z) 2 t_1^i \cosh \mu$.
For finite $N$, however, the bonding and antibonding states are formed by the surface states localized on the individual surfaces, leading to an energy gap $\Delta E$.
As an example, Figs. \ref{Fig_tb}(a) and (b) show the band structure of a 15-layer slab with
$t_p = -0.3$ and $\varepsilon = -2.5$ in units of $t_s$.
Two purely $p_z$-derived surface states appear below the bulk band continuum at $\bar{\Gamma}$, whose wave functions are localized on both surfaces, as shown in Fig. \ref{Fig_tb}(c).

Let us estimate $\Delta E$ by constructing the normalized bonding and antibonding states as
\begin{gather}
	\boldsymbol{c}_{\pm} = \frac{  \boldsymbol{c}_{\mathrm{L}} \pm \boldsymbol{c}_{\mathrm{R}}  }{ \sqrt{ 2 ( 1 \pm S ) }}
	, \quad
	S \equiv N e^{-\mu (N-1)}.
\end{gather}
We calculate the gap as a difference in the expectation values of the Hamiltonian between them and obtain
\begin{gather}
	\Delta E = {}^{\mathrm{t}} \boldsymbol{c}_- H^i \boldsymbol{c}_-
	- {}^{\mathrm{t}} \boldsymbol{c}_+ H^i \boldsymbol{c}_+
	= \frac{2}{1-S^2} (ES - {}^{\mathrm{t}} \boldsymbol{c}_{\mathrm{L}} H^i \boldsymbol{c}_{\mathrm{R}})
	.
	\label{delta_e}
\end{gather}
Since ${}^{\mathrm{t}} \boldsymbol{c}_{\mathrm{L}} H^i \boldsymbol{c}_{\mathrm{R}}$ is expected to contain contributions mainly from several sites around the midpoint of the chain,
this term is on the order of $e^{-\mu (N-1)}$, while $ES$ is obviously on the order of $N e^{-\mu (N-1)}$.
Equation (\ref{delta_e}) can thus be reduced to
\begin{gather}
	\Delta E 
	\approx 2 E N e^{-\mu (N-1)}
	\label{ediff}
	,
\end{gather}
which is consistent with the rapid decay of the gap at $\bar{\Gamma}$ obtained in the first-principles calculation (see Fig. \ref{Fig_band2}(c)).

\subsection{Bilayer model}

\begin{figure}[htbp]
\begin{center}
\includegraphics[keepaspectratio,height=9cm]{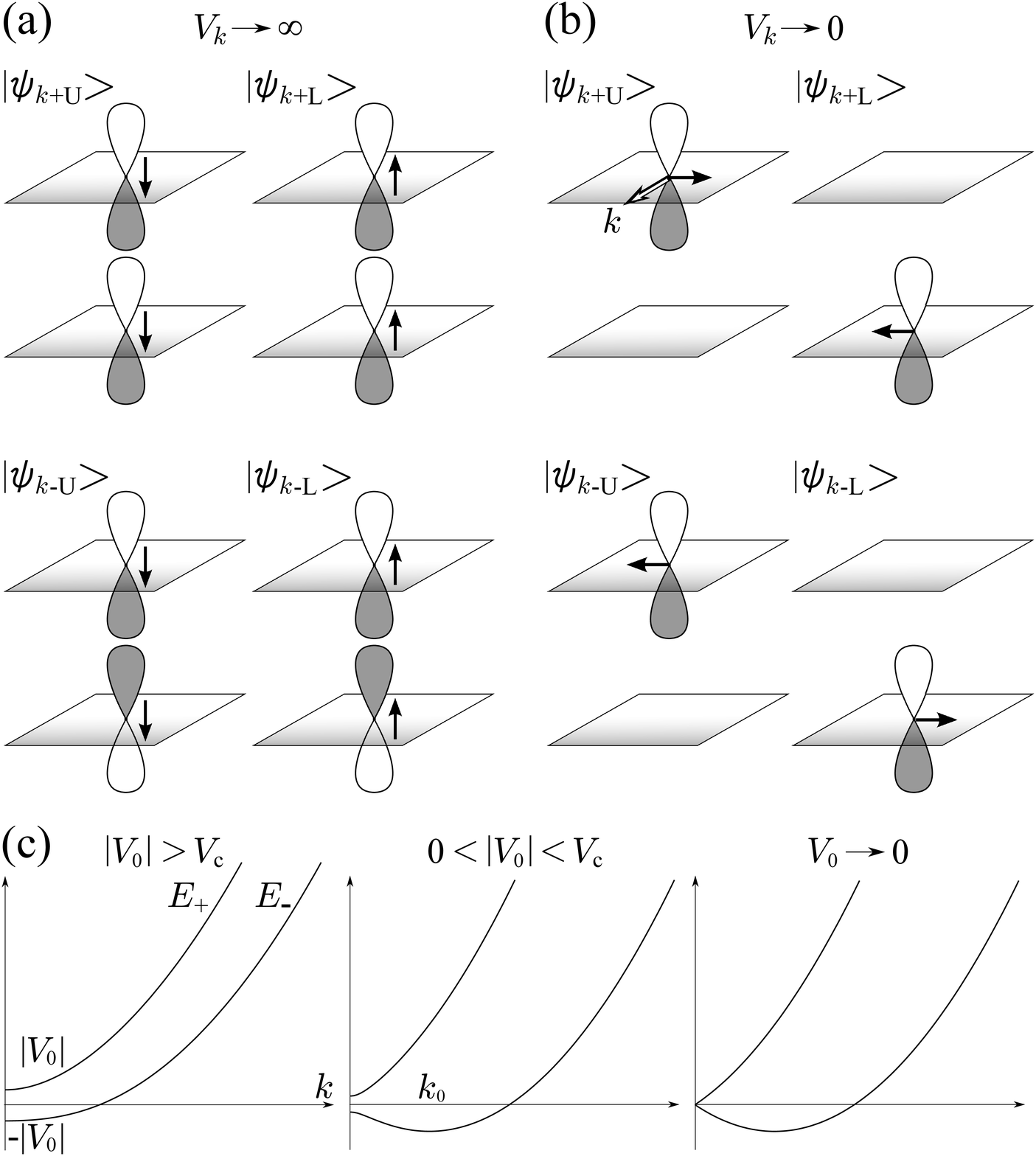}
\end{center}
\caption{
(a) and (b) are schematic illustrations of a bilayer two-dimensional free-electron model
in $V_{\boldsymbol{k}} \to \infty$ and $V_{\boldsymbol{k}} \to 0$ limits, respectively,
drawing the wave functions of the four eigenstates.
Filled arrows represent spin polarization.
(c) shows band structures of the bilayer system with $V_0$ varied.
}
\label{Fig_2dimfe}
\end{figure}

We here propose a bilayer two-dimensional free-electron model that continuously connects the two extreme situations,
in which either the interference of the surface states or the spin-orbit interaction is so strong that the other is negligible.
This model consists only of two layers to each of which free electrons are confined, feeling the spin-orbit interaction and the interference from the other layer.
The individual layers are described by the Rashba Hamiltonian, eq. (\ref{Rashba_H}).
The Hilbert space for this model is spanned by four states, namely,
free-electron-like wave functions on the upper and lower layers with up and down spins: $\{ |\mathrm{U} \uparrow \rangle, |\mathrm{U} \downarrow \rangle, |\mathrm{L} \uparrow \rangle, |\mathrm{L} \downarrow \rangle \}$.
The Hamiltonian matrix in $\boldsymbol{k}$ space reads
\begin{gather}
	H(\boldsymbol{k}) =
	\begin{pmatrix}
		\frac{ k^2}{2m} & -\alpha (i k_x + k_y) & V_{\boldsymbol{k}} e^{i \phi_{\boldsymbol{k}}} & 0 \\
		\alpha (i k_x - k_y) & \frac{ k^2}{2m} & 0 & V_{\boldsymbol{k}} e^{i \phi_{\boldsymbol{k}}}  \\
		V_{\boldsymbol{k}} e^{-i \phi_{\boldsymbol{k}}} & 0 & \frac{ k^2}{2m} & \alpha (i k_x + k_y)  \\
		0 & V_{\boldsymbol{k}} e^{-i \phi_{\boldsymbol{k}}} & -\alpha (i k_x - k_y) & \frac{ k^2}{2m} \\
	\end{pmatrix}
	,
\end{gather}
where $m$ is the effective mass of an electron confined to the layer.
The component $V_{\boldsymbol{k}} e^{i \phi_{\boldsymbol{k}}}$ ($V_{\boldsymbol{k}}$ and $\phi_{\boldsymbol{k}}$ are real) is responsible for the interference between the surface states on the confronting layers.
This matrix has two eigenvalues, to be compared with eq. (\ref{eigenvalues_monol}),
\begin{gather}
	E_\pm (\boldsymbol{k}) = \frac{ k^2}{2m} \pm \sqrt{V_{\boldsymbol{k}}^2 + \alpha^2 k^2},
	\label{eigenvalues_bil}
\end{gather}
each of which is twofold degenerate regardless of the $\boldsymbol{k}$ dependence of the interference component.
Considering the basic knowledge of the matrix theory that eigenvectors belonging to the same eigenvalue can be freely rotated by an arbitrary unitary matrix,
we adopt the following orthonormalized eigenvectors of $H(\boldsymbol{k})$:
\begin{gather}
	| \psi_{\boldsymbol{k} \pm \mathrm{U}} \rangle = \frac{1}{\sqrt{2}}
	\begin{pmatrix}
		\mp \frac{i  (k_x - i k_y) d_{\boldsymbol{k}} }{\sqrt{1 + k^2 d_{\boldsymbol{k}}^2}} \\
		1 \\
		0 \\
		\pm \frac{e^{-i \phi_{\boldsymbol{k}}} }{\sqrt{1 + k^2 d_{\boldsymbol{k}}^2}} 
	\end{pmatrix} ,
	| \psi_{\boldsymbol{k} \pm \mathrm{L}} \rangle = \frac{1}{\sqrt{2}}
	\begin{pmatrix}
		\pm \frac{e^{i \phi_{\boldsymbol{k}}} }{\sqrt{1 + k^2 d_{\boldsymbol{k}}^2 }} \\
		0 \\
		1 \\
		\mp \frac{i  (k_x + i k_y) d_{\boldsymbol{k}}}{\sqrt{1 + k^2 d_{\boldsymbol{k}}^2 }}
	\end{pmatrix}
	,
\end{gather}
where $d_{\boldsymbol{k}} \equiv \alpha/V_{\boldsymbol{k}}$ measures the strength of the spin-orbit interaction compared with the interference between the layers.
$| \psi_{\boldsymbol{k} + (-) \mathrm{U}} \rangle$ and 
$| \psi_{\boldsymbol{k} + (-) \mathrm{L}} \rangle$ belong to $E_{ + (-)} (\boldsymbol{k})$.
The degenerate eigenvectors for the same branch have opposite expectation values of the spin operator
$\langle \boldsymbol{S} \rangle_{\boldsymbol{k} \pm \mathrm{U}} = - \langle  \boldsymbol{S} \rangle_{\boldsymbol{k} \pm \mathrm{L}}
= (\mp k_y d_{\boldsymbol{k}}, \pm k_x d_{\boldsymbol{k}}, -1)/2 \sqrt{1 + k^2 d_{\boldsymbol{k}}^2 } $,
which is natural because of the inversion symmetry of the system.
The ratios of the weights on the layers of the eigenvectors are given by
\begin{gather}
	\frac{ \langle P_{\mathrm{U}} \rangle_{\boldsymbol{k} \pm \mathrm{U}} }
		{ \langle P_{\mathrm{L}} \rangle_{\boldsymbol{k} \pm \mathrm{U}} }
	=
	\frac{ \langle P_{\mathrm{L}} \rangle_{\boldsymbol{k} \pm \mathrm{L}} }
		{ \langle P_{\mathrm{U}} \rangle_{\boldsymbol{k} \pm \mathrm{L}} }
	= 1 + 2 k^2 d_{\boldsymbol{k}}^2
	,
	\label{weights}
\end{gather}
where $P_{\mathrm{U}}$ and $P_{\mathrm{L}}$ are the projection operators onto the upper and lower layers, respectively.
Equation (\ref{weights}) implies that, for $d_{\boldsymbol{k}} \ne 0$,
$| \psi_{\boldsymbol{k} \pm \mathrm{U}} \rangle$ and $| \psi_{\boldsymbol{k} \pm \mathrm{L}} \rangle$ have major components on the upper and lower layers, respectively.
Let us verify that $d_{\boldsymbol{k}}$ continuously connects the two limits, the strong interference limit and the strong spin-orbit interaction limit.
In the former limit, that is $d_{\boldsymbol{k}} \to 0$, the eigenvectors are
$| \psi_{\boldsymbol{k} \pm \mathrm{U}} \rangle = {}^{\mathrm{t}} ( 0, 1, 0, \pm e^{-i \phi_{\boldsymbol{k}}} ) /\sqrt{2}$ and
$| \psi_{\boldsymbol{k} \pm \mathrm{L}} \rangle = {}^{\mathrm{t}} ( \pm e^{i \phi_{\boldsymbol{k}}} , 0, 1, 0 ) /\sqrt{2}$.
$| \psi_{\boldsymbol{k} + (-) \mathrm{U}} \rangle$ and 
$| \psi_{\boldsymbol{k} + (-) \mathrm{L}} \rangle$ 
are antibonding (bonding) states with equal weights on both layers with purely down and up spins, respectively (see Fig. \ref{Fig_2dimfe}(a)).
In the latter limit, on the other hand, that is $d_{\boldsymbol{k}} \to \infty$, the eigenvectors are
$| \psi_{\boldsymbol{k} \pm \mathrm{U}} \rangle = {}^{\mathrm{t}} (\mp i (k_x - i k_y)/k, 1, 0, 0 ) /\sqrt{2} $ and 
$| \psi_{\boldsymbol{k} \pm \mathrm{L}} \rangle = {}^{\mathrm{t}} (0, 0, 1, \mp i (k_x + i k_y)/k ) /\sqrt{2}$,
for which the two layers are decoupled and each of these layers becomes an ordinary two-dimensional free-electron system with the Rashba term (see Fig. \ref{Fig_2dimfe}(b)).

We then examine the dispersion relation of the bilayer system,
regarding the interference component to be real and $\boldsymbol{k}$-independent, $V_0$, for easier qualitative understanding.
The eigenvalues, eq. (\ref{eigenvalues_bil}), take extrema at $k = 0$ when $V_0 \ne 0$,
giving an $\alpha$-independent energy gap of $E_+(0) - E_-(0) = 2 |V_0|$.
In addition, if $|V_0| < m \alpha^2  \equiv V_{\mathrm{c}}$, $E_-$ takes a minimum at
$k_0 \equiv \sqrt{V_{\mathrm{c}}^2 - V_0^2}/\alpha$ and thereby the gap is
$E_+(k_0) - E_-(k_0) = 2 V_{\mathrm{c}}$, independent of $V_0$ (see Fig. \ref{Fig_2dimfe}(c)).
The $V_0$-independence of this gap is consistent with the result of the first-principles calculation, shown in Fig. \ref{Fig_band2}(c).
The ratio $V_{\mathrm{c}}/|V_0|$ characterizes the crossover of the band splitting from the bonding-dominant nature to the spin-dominant nature.
Combining the result of eq. (\ref{ediff}) for the tight-binding model analysis,
we can obtain the condition for the number of layers in order for the spin-split bands to be reproduced in a slab calculation.
The gap at $k = 0$ should be identified with $\Delta E = 2 c_1 N c_2^{-N}$, where $c_1$ and $c_2$ is independent of $N$,
and hence the condition for nonzero $k_0$ is
\begin{gather}
	c_1 N c_2^{-N} < V_{\mathrm{c}}
	.
\end{gather}
This is a necessary condition for reproducing the spin-split band structure characteristic to the ideal Rashba system, described by the Rashba Hamiltonian, eq. (\ref{Rashba_H}).
It tells us explicitly that a slab consisting of sufficiently many layers allows the Rashba splitting to be reproduced despite the presence of the inversion symmetry in a practical electronic structure calculation,
that is, the smallest $N$ that satisfies this condition is $15$ in our first-principles calculation for a Au(111) slab.

\section{Summary}

We performed fully relativistic first-principles calculations of the electronic structure of Au(111) slabs with varying the number $N$ of atomic layers and the strength $\lambda$ of spin-orbit interaction.
The variation of the features of the bands of the $sp$-derived surface states in the vicinity of $\bar{\Gamma}$ was examined with either one of $N$ and $\lambda$ varied and the other one fixed.
We found, for $N \geq 15$, that the finite wave vector $k_0$ exists, for which the energy gap is nearly unchanged when $N$ is increased, in contrast to the gap at $\bar{\Gamma}$.
The rapid decay of the gap at $\bar{\Gamma}$ as a function of $N$ was explained by the simple tight-binding calculation of a slab including only $p$ orbitals.
We adopted the bilayer two-dimensional free-electron model
and were able to clearly understand the qualitative behavior of the first-principles band structure with $N$ and $\lambda$ varied.
In addition, combining the results of the tight-binding and bilayer models, we obtained the explicit condition of $N$ and $\alpha$ for occurrence of spin-split bands in an electronic structure calculation using a slab.
From the success of the bilayer model,
the combination of first-principles calculations and simplified low-dimensional models is expected to help one to understant 
real systems more complicated than a slab, such as an overlayer on a substrate and an interface.
Furthermore, the insights obtained in the present work are useful for achieving a reliable electronic structure calculation of a surface system
because they help one to distinguish artifacts coming from the finiteness of a slab and the intrinsic properties of the surface system.

\section*{Acknowledgements}
The authors are grateful to Professors
Tamio Oguchi, Taisuke Ozaki, and Kiyoyuki Terakura for fruitful discussions. 
The present work is partially supported by the Next Generation Supercomputer Project,
Nanoscience Program from MEXT, Japan, and 
by KAKENHI under Grant No. 22104010 from MEXT, Japan.
The calculations were performed at the supercomputer centers of ISSP, 
University of Tokyo, and at the Information Technology Center, University of Tokyo.

\end{document}